\documentclass[sigconf]{acmart}
\usepackage{graphicx}
\usepackage{subcaption}
\usepackage{multirow}
\usepackage{makecell}
\usepackage{algorithm}
\usepackage{algpseudocode}

\AtBeginDocument{%
  \providecommand\BibTeX{{%
    \normalfont B\kern-0.5em{\scshape i\kern-0.25em b}\kern-0.8em\TeX}}}

\setcopyright{acmlicensed}
\copyrightyear{2018}
\acmYear{2018}
\acmDOI{XXXXXXX.XXXXXXX}

\acmISBN{978-1-4503-XXXX-X/18/06}

\begin{document}

\title{CoRAL: Collaborative Retrieval-Augmented Large Language Models Improve Long-tail Recommendation}

\author{Junda Wu}
\email{juw069@ucsd.edu}
\affiliation{%
  \institution{University of California San Diego}
  \city{La Jolla}
  \state{California}
  \country{USA}
}

\author{Cheng-Chun Chang}
\email{cc4900@columbia.edu}
\affiliation{%
  \institution{Columbia University}
  \city{New York}
  \state{New York}
  \country{USA}
}

\author{Tong Yu}
\email{tyu@adobe.com}
\affiliation{%
  \institution{Adobe Research}
  \city{San Jose}
  \state{California}
  \country{USA}
}

\author{Zhankui He}
\email{zhh004@eng.ucsd.edu}
\affiliation{%
  \institution{University of California San Diego}
  \city{La Jolla}
  \state{California}
  \country{USA}
}

\author{Jianing Wang}
\email{lygwjn@gmail.com}
\affiliation{%
  \institution{University of California San Diego}
  \city{La Jolla}
  \state{California}
  \country{USA}
}
% option:
% \affiliation{%
%   \institution{East China Normal University}
%   \city{Shanghai}
%   \state{}
%   \country{China}
% }

\author{Yupeng Hou}
\email{yphou@ucsd.edu}
\affiliation{%
  \institution{University of California San Diego}
  \city{La Jolla}
  \state{California}
  \country{USA}
}

\author{Julian McAuley}
\email{jmcauley@ucsd.edu}
\affiliation{%
  \institution{University of California San Diego}
  \city{La Jolla}
  \state{California}
  \country{USA}
}

\renewcommand{\shortauthors}{Wu, et al.}

\begin{abstract}
The long-tail recommendation is a challenging task for traditional recommender systems, due to data sparsity and data imbalance issues.
The recent development of large language models (LLMs) has shown their abilities in complex reasoning, which can help to deduce users' preferences based on very few previous interactions.
However, since most LLM-based systems rely on items' semantic meaning as the sole evidence for reasoning, the collaborative information of user-item interactions is neglected, which can cause the LLM's reasoning to be misaligned with task-specific collaborative information of the dataset.
To further align LLMs' reasoning to task-specific user-item interaction knowledge, we introduce collaborative retrieval-augmented LLMs, \textbf{CoRAL}, which directly incorporate collaborative evidence into the prompts.
Based on the retrieved user-item interactions, the LLM can analyze shared and distinct preferences among users, and summarize the patterns indicating which types of users would be attracted by certain items.
The retrieved collaborative evidence prompts the LLM to align its reasoning with the user-item interaction patterns in the dataset.
However, since the capacity of the input prompt is limited, finding the minimally-sufficient collaborative information for recommendation tasks can be challenging.
We propose to find the optimal interaction set through a sequential decision-making process and develop a retrieval policy learned through a reinforcement learning (RL) framework, \textbf{CoRAL}.
Our experimental results show that CoRAL can significantly improve LLMs' reasoning abilities on specific recommendation tasks. 
Our analysis also reveals that CoRAL can more efficiently explore collaborative information through reinforcement learning.

\end{abstract}





\maketitle

\section{Introduction}

Recommendation systems are valuable tools for users to explore content that matches their preferences. 
Traditional data-driven recommendation algorithms (\emph{e.g.} , collaborative filtering) 
can fail in long-tail recommendation, due to the uneven distribution in user-item interactions \cite{liu2020long,zhang2023empowering,luo2023improving,rahmani2022experiments}.
In this paper, we aim to address the challenges associated with long-tail items in collaborative filtering-based recommender systems \cite{zhang2023empowering,zhang2021model}. 
In such scenario, long-tail items may have very few associated interactions, such that data-driven algorithms cannot accurately capture user-item interaction patterns \cite{gong2023full,liu2023co,tang2021cadpp}.
In addition, models trained on such uneven datasets can suffer from selection bias \cite{ovaisi2020correcting,wang2016learning}, exposure bias \cite{gupta2021correcting,khenissi2020modeling} and popularity bias \cite{wei2021model,zhang2023model,abdollahpouri2021user}. These biases can cause the models to overfit on popular items.

To tackle popularity bias and improve long-tail recommendation performance, data augmentation, and re-balancing methods can be directly applied.
Data re-balancing methods \cite{menon2020long,yi2019sampling,byrd2019effect,cui2019class} try to reduce the distribution discrepancy between popular items and long-tail items in the training stage. 
However, these methods often obtain sub-optimal solutions due to learning inefficiency problems on long-tail data \cite{zhang2023empowering, zhang2021model}. 
This inefficiency leads to knowledge forgetting in the majority of the data, namely the popular items \cite{zhang2023empowering}.
Since the goal of these methods is to achieve a compromise between the model's attention on popular items and more diversified recommendations on long-tail items, 
achieving accurate recommendations for long-tail items can be challenging.
Causal debiasing learning \cite{zheng2021disentangling,schnabel2016recommendations,bonner2018causal} is another line of work that focuses on how to learn the underlying user preferences, 
instead of simply learning user-item correlation from the data. 
Since long-tail items only have limited numbers of previous interactions, the model's fine-grained reasoning abilities become essential for learning user preferences.

\begin{figure}[t]
    \begin{subfigure}[b]{0.48\textwidth}
        \centering
        \includegraphics[width=\textwidth]{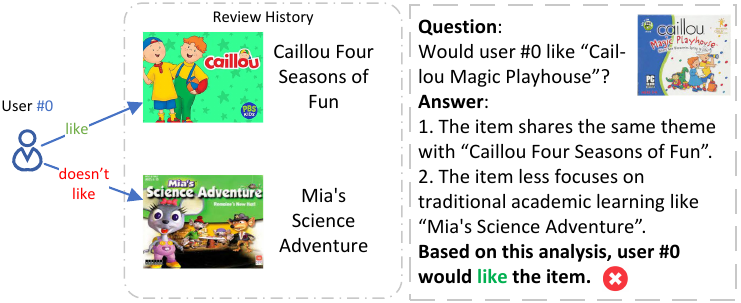}
        \caption{Conventional item-based \cite{sanner2023large, harte2023leveraging} LLM reasoning process.}
        \label{fig:intro1}
    \end{subfigure}%
    \hfill % Adds horizontal space between subfigures
    \begin{subfigure}[b]{0.48\textwidth}
        \centering
        \includegraphics[width=\textwidth]{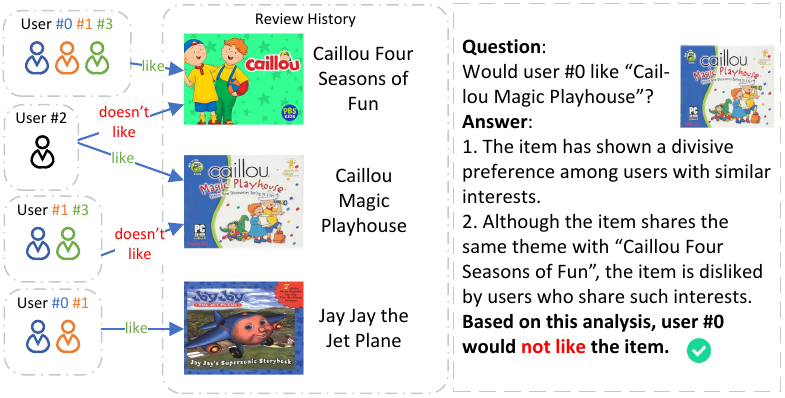}
        \caption{Collaborative Retrieval Augmented LLM reasoning process.}
        \label{fig:intro2}
    \end{subfigure}

    \caption{
    By text comprehension and extracting information-rich semantic features \cite{runfeng2023lkpnr},
    LLMs in (a) can handle long-tail items \cite{sanner2023large}, but still cannot directly leverage collaborative information.
    To handle long-tail items in collaborative filtering-based recommender systems, 
    by collaborative prompting, LLMs in (b)
    can reason the fact that even if the current item shares the same theme with previously liked items,
    users with similar interests still dislike this item, which provides the rationale to not recommending it.
    }
    \label{fig:intro}
    % \vspace{-.8cm}
\end{figure}

Large language models (LLMs) have recently demonstrated great reasoning abilities on very complex reasoning tasks \cite{tan2023can,yu2023thought,wang2024instructgraph}, 
in which fine-grained reasoning paths can be generated to help with obtaining the correct answers \cite{wei2022chain,zhang2022automatic}. 
Previous works have also tried to adapt LLMs' reasoning abilities to recommender systems \cite{wang2023drdt,wang2023recmind}.
One line of previous works tries to use the language description of item content as the reasoning context \cite{sanner2023large, harte2023leveraging}, 
which can be augmented by the LLM's internal knowledge \cite{zhang2023bridging,yao2023knowledge,wei2023llmrec,baek2023knowledge}.
By representing items as natural language, item representation distribution is aligned with the LLM's knowledge.
This alignment allows for a universal semantic representation of items, potentially mitigating the issue of long-tail items.
In addition, by aligning recommendation tasks to the reasoning paradigms of language models, 
LLMs can be leveraged for more fine-grained reasoning based on the semantic contexts of users and items \cite{wang2023drdt,wang2023recmind}. 
However, due to several misalignment issues of LLMs in recommendation \cite{ma2023large}, directly prompting LLMs can be problematic.
Specifically, the LLM's understanding of user preferences over items can be misaligned with real user-item interaction patterns.
For example, in Figure~\ref{fig:intro1}, conventional LLM-based methods may simply recommend similar items (\emph{e.g.}, ``Caillou Magic Playhouse'' is recommended because the user likes ``Caillou Four Seasons of Fun'').

In this work, we propose to formulate long-tail recommendation tasks as natural language reasoning problems and use LLMs to enable fine-grained reasoning about user preferences on long-tail items.
To further align the reasoning of LLMs with task-specific knowledge of user-item interactions, 
we introduce collaborative retrieval-augmented LLMs, \textbf{CoRAL}, which directly incorporate collaborative evidence into the prompts, via collaborative prompting.
For example, in Figure~\ref{fig:intro2}, additional user-item interaction information can be retrieved by an additional lightweight model and included in the prompt. 
Based on the retrieved collaborative information, the LLM can find that although the items share the same theme (\emph{e.g.}, ``Caillou''), the item (\emph{e.g.}, ``Caillou Four Seasons of Fun'') is disliked
by users who share such interests (\emph{e.g.}, preference on ``Caillou Four Seasons of Fun'').
However, due to the limited size of the prompts, a large amount of collaborative information cannot be included, and duplicate information may also distract the LLM's reasoning process.
Thus, we develop a retrieval policy to find the minimal-sufficient collaborative information which serves as the supporting evidence of the LLM's reasoning on the given user-item pair.
Since the number of users and items in a recommendation task is significantly larger than the capacity of a prompt input in LLMs,
the retrieval policy should learn to explore diversified users and items for potential information gain,
as well as exploit the collected collaborative information to maximize prediction accuracy.
Based on the necessity to balance between exploration and exploitation in learning the retrieval policy, 
we propose to formulate the retrieval process as a sequential decision-making problem and employ reinforcement learning methods to maximize the long-term reward.

To improve the data efficiency and reduce the early exploration time, we also propose to use the data from popular items to provide a warm start for the learning of the retrieval policy.
Before the reinforcement learning stage, the user and item representations are learned from the data from popular items by conventional collaborative filtering methods.
Then, the retrieval policy network will be initialized by the learned user and item representations, which improves the exploration efficiency and helps to solve the reward-sparsity problem.
We summarize our contributions as follows:
\begin{itemize}
    \item We identify the misalignment problem between the LLM's reasoning process and long-tail recommendation tasks, which is caused by the lack of collaborative information. 
    \item We propose to retrieve additional user-item interactions as collaborative information for collaborative prompting, which helps to align the LLM's reasoning process to general recommendation tasks. 
    \item We formulate the retrieval process as a sequential decision-making task and propose an RL framework, 
    in which the retrieval policy learns to find the minimal-sufficient collaborative information specific to a recommendation task. 
    \item To further improve data efficiency, we propose to learn the prior knowledge from more abundant data of popular items and to provide a warm start for the retrieval policy.
\end{itemize}
\section{Related Works}
\subsection{Long-tail Recommendation}
Long-tail recommendation plays a crucial role in mitigating the issue of highly skewed distributions of long-tail items in recommendation tasks. 
Previous works have proposed knowledge transfer learning methods based on novel model architecture designs.
\citet{zhang2021model} introduces a dual transfer learning framework by leveraging a model-level knowledge transfer and an item-level transfer to link head and tail items through shared features. 
\citet{zhang2023empowering} propose a Cross Decoupling Network (CDN). This network aims to improve tail item recommendations while simultaneously maintaining overall system efficiency.
\citet{wu2022dynamics} propose a domain transfer learning method (DACIR) for the sequential recommendation.
However, the popularity bias in such long-tail or cold-start recommendation datasets is less discussed in their model designs.
To address the issue of bias within the dataset in recommender systems, 
extensive efforts \cite{zheng2021disentangling,liu2024interact,bonner2018causal,xia2023user,wu2021deconfounded} have been dedicated to designing different training frameworks via causal debiasing.
In this work, we propose to handle long-tail items by leveraging LLMs' abilities in fine-grained reasoning on collaborative information and extracting rich semantic features. 

\subsection{LLMs in Recommender Systems}
A few studies underscore the expanding role of LLMs in recommender systems.  
\citet{harte2023leveraging} focus on enhancing sequential recommendation models with LLM embeddings, while \citet{sanner2023large} explore the use of LLMs in processing language-based user preferences in dialog interfaces. 
A surge of approaches \cite{zhang2023bridging,yao2023knowledge,wei2023llmrec,baek2023knowledge} propose content augmentation method to reduce the cost of re-training or fine-tuning LLMs.  
To improve LLMs' reasoning capability for recommender systems, \cite{wang2023recmind} proposes RecMind, specifically designed to deliver personalized recommendations. 
\citet{wang2023drdt}, on the other hand, use a retriever-reranking framework to enhance collaborative in-context understanding. 
However, the LLM's understanding of user preferences over items can be misaligned with real user-item interaction patterns, due to the lack of enough collaborative information.

To further align the LLM's reasoning process to specific recommendation tasks, several works have proposed to inject collaborative knowledge into LLMs by soft-prompt instruction tuning \cite{zhang2023collm,zheng2023adapting}.
\cite{li2023prompt} also introduces a method to transform discrete task-specific prompts into continuous prompt vectors, effectively linking IDs and words while decreasing inference time.
However, such methods require an even larger amount of data to achieve good alignment, which would not help in our long-tail recommendation setting. 
Instead, we propose a lightweight retrieval policy to augment collaborative information in LLMs' reasoning process.

\section{Problem Formulation}
In this paper, we focus on collaborative filtering-based recommender systems with long-tail items \cite{zhang2023empowering,zhang2021model}.
We formulate long-tail recommendation as a complex reasoning task in LLMs \cite{wang2023boosting,li2023theory,kim2023fineprompt}. 
When the LLM-empowered recommender system interacts with a user $u\in \mathcal{U}$, the task is to predict the user's preference for a long-tail item $i\in \mathcal{I}$.
Since LLMs have no prior internal knowledge about a certain user's preference as well as the collaborative information of a certain recommendation task, the reasoning process can only be enabled by incorporating supporting evidence.
To provide the information for the user $u$, items $\mathcal{I}^{\mathit{supp}}_u=(i^u_1,i^u_2,\dots,i^u_m)\subseteq \mathcal{I}$ from the user's previous interactions are included, 
which helps the LLM to understand the preference of the user \cite{wang2023recmind,kang2023llms,bao2023tallrec}.
To help the LLM further understand how the user $u$ would rate a certain item $i$, collaborative information will be retrieved as evidence of user-item interaction patterns.
Due to the limitation of the LLM's reasoning context capacity, instead of including all the user-item interaction information, for a certain user-item pair $z =(u, i)$, the retrieval policy $\pi_{\theta}$ will find 
a sequence of supporting users $\mathcal{U}^{\mathit{coll}}_{z}=(u^{z}_1,u^{z}_2,\dots,u^{z}_t)\subseteq \mathcal{U}$ and a sequence of supporting items  $\mathcal{I}^{\mathit{coll}}_{z}=(i^z_1,i^z_2,\dots,i^z_t)\subseteq \mathcal{I}$.
At each time step $t$, the retrieval policy $\pi_{\theta}$ needs to retrieve the next user-item pair $(u^z_{t+1},i^z_{t+1})$ to augment current supporting evidence.
In this work, we focus on how to obtain a minimal-sufficient information support for the LLM to deduce the accurate rating of $z$.

\subsection{MDP Formulation for Retrieval Policy}
We formulate the sequential retrieval process as a Markov Decision Process (MDP) $\mathcal{M}=(\mathcal{S}, \mathcal{A}, P, r, \rho, \gamma)$, where
\begin{itemize}
    \item $\mathcal{S}$ is a continuous state space that encodes the collaborative information from the retrieved users and items as well as their collaborative preference patterns. The details of the design of state $\boldsymbol{s}\in \mathcal{S}$ encoding networks are explained in Section \ref{sec:policy}.
    \item $\mathcal{A}$ is a continuous action space that represents the retrieval queries of the next user and next item.
    At time step $t+1$, the retrieval queries will try to retrieve the most relevant user and item in their feature spaces, as well as try to explore potentially useful users and items in the under-explored regions.
    The details of the action $\boldsymbol{a} \in \mathcal{A}$ prediction are explained in Section \ref{sec:policy}. 
    \item $P: \mathcal{S}\times\mathcal{A}\times\mathcal{S}\rightarrow \mathbb{R}$, is the state transition probability distribution, which captures the dynamics of the retrieval process.
    \item $r: \mathcal{S}\times\mathcal{A}\rightarrow \mathbb{R}$, is the reward function $r(\boldsymbol{s}, \boldsymbol{a})$ given current state $\boldsymbol{s}$ and action $\boldsymbol{a}$. The details of the reward function design are explained in Section \ref{sec:reward}.
\end{itemize}
At each time step $t$, the policy will generate the retrieval query $\boldsymbol{a}_t\in \mathcal{A}$ and retrieve the next user-item pair $(u^{z}_t,i^{z}_t)$.
The learning objective is to find the optimal retrieval policy $\pi^*_{\theta}:\mathcal{S}\rightarrow \mathcal{A}$ to achieve the long-term goal of obtaining a minimal-sufficient information support for the LLM, by maximizing the cumulative reward:
\begin{equation*}
    \pi^*_{\theta}=\arg \max _{\pi \in \Pi} \mathbb{E}\left[\sum_{t=0}^{T} \gamma^t r\left(\boldsymbol{s}_t, \boldsymbol{a}_t\right)\right],
\end{equation*}
in which $\gamma$ is the discount rate of future rewards and $\Pi$ is the policy search space. When the retrieval policy observes the next user-item pair $(u^{z}_t,i^{z}_t)$, the state is updated by encoding the user-item pair into the state information $\boldsymbol{s}_{t+1}=P(\cdot|\boldsymbol{s}_t,[u^{z}_t,i^{z}_t])$.

\subsection{Reward Function} \label{sec:reward}
For each time step $t$, the user-item rating prediction $y_t$ will be prompted from the large language model $P_\phi$ by the context $C_t$ constructed from user information $\mathcal{I}^{\mathit{supp}}_u$, and previously collected collaborative information $\mathcal{U}^{\mathit{coll}}_{z}$ and $\mathcal{I}^{\mathit{coll}}_{z}$,
\begin{align}
    p_t &= P_\phi(y_{t}|C_t), \label{eq:prob} \\ 
    C_t &= C\left(\mathcal{I}^{\mathit{supp}}_u, \mathcal{U}^{\mathit{coll}}_{z}, \mathcal{I}^{\mathit{coll}}_{z}\right), \label{eq:prp}
\end{align}
in which $p_t$ is the prediction likelihood of whether the user $u$ likes the item $i$, and $C$ is the prompt template (detailed description in Section \ref{sec:prompt}) which composes the collected information into a natural language query.

Since the motivation of the retrieval policy is to maximize cumulative information gain, we use the marginal information gain at each time step $t$ as the reward signal $r_t$, which is calculated by the prediction discrepancy,
\begin{equation} \label{eq:reward}
    r_t\left(s_t, (u^{z}_t,i^{z}_t)\right)=
    \underbrace{\left|p_{t-1}-y^{g t}\right|}_{\text {discrepancy at } t-1}-
    \underbrace{\left|p_t-y^{g t}\right|}_{\text {discrepancy at } t},
\end{equation}
in which $y^{gt}=\textbf{M}(u, i)$ is the ground truth label of the user's preference on the item from the training rating matrix $M$. 
Following \cite{zhuang2022spatial,murahari2019improving}, we reward those retrieved user-item pairs which will lead the constructed prompt to find a more accurate prediction based on the LLM $P_\phi$. 

\section{Proposed Framework: CoRAL}
In this section, we first explain the prompting method designed to incorporate collaborative information and collect the LLM's prediction as the recommendation prediction.
Then, we introduce the collaborative retrieval policy network as well as the reinforcement learning process, which is illustrated in Algorithm \ref{alg:ddpg}.
In reinforcement learning, we treat the LLM as part of the environment and thus the LLM is frozen while a lightweight retrieval policy is learnable with significantly fewer learning parameters. To further improve the policy's learning efficiency and accommodate long-tail recommendation, we propose to use collaborative filtering models learned on the short-head data as the model initialization (detailed experimental settings and comparison results are in Section \ref{sec:transfer}).

\subsection{Collaborative Prompting} \label{sec:prompt}
In this section, we will explain how to construct the prompt $C_t$ with the retrieved collaborative information in Eq. \eqref{eq:prp}, and how to obtain the prediction probability $p_t$ in Eq. \eqref{eq:prob}, given the retrieval results from the policy $\pi_\theta$.
Details about the policy network design will be explained in Section \ref{sec:policy}.

\textbf{Collaborative Information}. At  time step $t$, given the user-item pair $z=(u,i)$, the retrieval policy $\pi_\theta$ obtains the supporting users $\mathcal{U}^{\mathit{coll}}_{z}$ and items $\mathcal{I}^{\mathit{coll}}_{z}$.
To describe the users and items in natural language and incorporate them into the prompt, we represent each user $u^{z}_t\in \mathcal{U}^{\mathit{coll}}_{z}$ by their user index $\textbf{idx}_{\mathcal{U}}(u^{z}_t)$, 
and each item $i^z_t\in \mathcal{I}^{\mathit{coll}}_{z}$ by its item index $\textbf{idx}_{\mathcal{I}}(i^{z}_t)$. We further extract a short text description $\textbf{desc}_{\mathcal{I}}(i^{z}_t)$ for each item from the metadata (detailed descriptions in Section \ref{sec:data}), to assist the LLM's understanding of the item.
Based on the rating matrix $M$ in the training dataset, we can summarize users' shared preference for each item $i\in \mathcal{I}^{\mathit{coll}}_{z}$ in the following format:
\begin{align}
    \textbf{POS}(i, \mathcal{U}^{\mathit{coll}}_{z}) &= \left\{\textbf{M}(i,u)\geq y^{thresh}, u\in \mathcal{U}^{\mathit{coll}}_{z} \right\}, \nonumber \\ 
    \textbf{NEG}(i, \mathcal{U}^{\mathit{coll}}_{z}) &= \left\{\textbf{M}(i,u)<y^{thresh}, u\in \mathcal{U}^{\mathit{coll}}_{z} \right\}, \label{eq:collab}
\end{align}
in which the rating threshold $y^{thresh}$ is to determine if the rating is positive or negative. 
By aggregating the preference of a group of users for each item, the length of the prompt can be significantly reduced, and such descriptions prompt the LLM to focus more on the comparative preference among the users. 
To construct the first part of the prompt which contains collaborative information, we design the prompt as follows: 
\begin{itemize}
    \item \textbf{Role-play}:  As a recommender system please solve the following problem.
    \item \textbf{Collaborative Information}: Repeat $i \in\mathcal{I}^{\mathit{coll}}_{z}$
    \item[] $\left\{
    \begin{array}{l}
    \text {The item } \textbf{desc}_{\mathcal{I}}(i) \text { is liked by the users } \textbf{POS}(i, \mathcal{U}^{\mathit{coll}}_{z}).  \\ 
    \text {The item } \textbf{desc}_{\mathcal{I}}(i) \text { is disliked by the users } \textbf{NEG}(i, \mathcal{U}^{\mathit{coll}}_{z}). 
    \end{array}\right.$
    \item \textbf{Summarization}:  Try to understand the pattern that the item $\textbf{desc}_{\mathcal{I}}(i)$ is typically liked by what kinds of users based on the above information. 
\end{itemize}
Based on our empirical observations, the last \textbf{Summarization} instruction is essential to align the LLM's reasoning with the goal of this task.

\textbf{User Preference Representation}. To include more information on the user's preference, we follow previous works \cite{zhang2023bridging,yao2023knowledge,wei2023llmrec,baek2023knowledge} to include the user's previously interacted items $\mathcal{I}^{\mathit{supp}}_{z}$ and their text descriptions.
However, different from previous works, we also divide the previous items $\mathcal{I}^{\mathit{supp}}_{z}$ of user $u$ into positive and negative sets, and then query the LLM to deduce the rating for the user-item pair $z=(u,i)$,
\begin{align}
    \textbf{POS}(\mathcal{I}^{\mathit{supp}}_{z}, u) &= \left\{\textbf{M}(i,u)\geq y^{thresh}, i\in \mathcal{I}^{\mathit{coll}}_{z} \right\}, \nonumber \\
    \textbf{NEG}(\mathcal{I}^{\mathit{supp}}_{z}, u) &= \left\{\textbf{M}(i,u)<y^{thresh}, i\in \mathcal{I}^{\mathit{coll}}_{z} \right\}, \label{eq:user}
\end{align}
Then, we construct the second part of the prompt by including the user's previously interacted items $\mathcal{I}^{\mathit{supp}}_{z}$:
\begin{itemize}
    \item \textbf{User's Positive Preference}: Items the user $\textbf{idx}_{\mathcal{U}}(u)$ likes are as follows: $\textbf{POS}(\mathcal{I}^{\mathit{supp}}_{z}, u)$.
    \item \textbf{User's Negative Preference}: Items the user $\textbf{idx}_{\mathcal{U}}(u)$ does not likes are as follows:  $\textbf{NEG}(\mathcal{I}^{\mathit{supp}}_{z}, u)$.
    \item \textbf{Query}:  For the item described as $\textbf{idx}_{\mathcal{I}}(i)$, would you recommend it to the user $\textbf{idx}_{\mathcal{U}}(u)$? 
\end{itemize}

With the prompt design (denoted as $C$) described above, we can aggregate the information retrieved at time step $t$ and transform the information into a natural language prompt 
$C_t = C\left(\mathcal{I}^{\mathit{supp}}_u, \mathcal{U}^{\mathit{coll}}_{z}, \mathcal{I}^{\mathit{coll}}_{z}\right)$.
To obtain the LLM's prediction as well as its confidence score, we extract the prediction probability $p_t = P_\phi(y_{t}|C_t)$ of the next token generated from the LLM.
Specifically, we strictly ask the LLM to answer either ``Yes'' or ``No'' without additional text provided, and we take the probability of the LLM generating the token ``Yes'' as our final score $p_t$.

\subsection{Retrieval Policy Network} \label{sec:policy}
In this section, we design the retrieval policy $\pi_\theta$ to sequentially include additional users and items, which may provide an information gain for the LLM's reasoning. 
Since the prompt has only a limited capacity of users and items included, the goal of the retrieval policy is to 
construct a minimal-sufficient prompt that contains complete information about the current recommendation task of the user-item pair $z=(u,i)$, 
Specifically, the retrieval policy needs to maximize its long-term information gain by maximizing the cumulative reward function.

Instead of learning the action distribution over all the users and items like value-based reinforcement learning methods \cite{mnih2013playing,schulman2017proximal}, 
we choose to directly learn the continuous vector representations of the next user and item based on the DDPG algorithm \cite{lillicrap2015continuous},
which helps to learn a low-rank decision space and also makes the solution more scalable even with new users and items included during the inference stage.

\subsubsection{State Representation}
For each user-item pair $z=(u,i)$, the retrieval process starts with the user-item embedding $\boldsymbol{s}_0=[\boldsymbol{u},\boldsymbol{i}]\in \mathbb{R}^{2d}$, in which $\boldsymbol{u}$ and $\boldsymbol{i}$ are the user and item embeddings in $d$ dimensions.
Notably, the user and item embeddings are randomly initialized by the multivariate normal distribution $\boldsymbol{u} \sim \mathcal{N}(\boldsymbol{\mu},\boldsymbol{\Sigma})$ and $\boldsymbol{i} \sim \mathcal{N}(\boldsymbol{\mu},\boldsymbol{\Sigma})$, 
in which $\boldsymbol{\mu}$ is a $d$-dimensional zero-vector and $\boldsymbol{\Sigma}$ is a $d\times d$ unit matrix. 

During the early stage of the reinforcement learning process, when the retrieval policy behaves randomly, 
similar users and items are likely to be retrieved due to the large-scale user and item spaces, which makes the reward of the policy's exploration very sparse. 
To overcome the exploration difficulty, we initialize the policy with the embeddings pre-trained on the portion of the dataset with popular items, 
which can provide a warm start for the learning of the retrieval policy (detailed comparison results are explained in Section \ref{sec:exp-main} and Section \ref{sec:transfer}).

\subsubsection{User-item Retrieval}
At each time step $t$, based on the current state $\boldsymbol{s}_t$, the retrieval policy $\pi_\theta$ will 
%act to 
find the next user-item pair. Due to the large user and item spaces, direct exploration in the discrete spaces of the users and items can be extremely inefficient.
Thus, we employ a continuous action space $\mathcal{A}\subseteq \mathbb{R}^{2d}$ which also covers the user-item embedding space. The retrieval policy will first generate a user-item query based on the current state $\boldsymbol{a}_{t+1} = [\boldsymbol{a}_{t+1}^u, \boldsymbol{a}_{t+1}^i] = \pi_\theta \left(\cdot|\boldsymbol{s}_t \right)$,
and try to find the nearest user and item in terms of a distance measurement $d(\cdot,\cdot)$ defined on the embedding spaces,
\begin{equation} \label{eq:action}
    u^{z}_{t+1} = \min_{u\in \mathcal{U}} d(\boldsymbol{u},\boldsymbol{a}_{t+1}^u),  \quad
    i^{z}_{t+1} = \min_{i\in \mathcal{U}} d(\boldsymbol{i},\boldsymbol{a}_{t+1}^i),  
\end{equation}
in which $\boldsymbol{u}$ and $\boldsymbol{i}$ denote the embeddings of the user $u$ and item $i$ respectively, and we use Euclidean distance for $d$.
The retrieved user and item will be added to the collaborative information $\mathcal{U}^{\mathit{coll}}_{z}$ and $\mathcal{I}^{\mathit{coll}}_{z}$.

\subsubsection{State Transition}
The state $\boldsymbol{s}_t$ encodes the current collaborative information, which will be updated for each time step after the user and item is retrieved. 
To track the retrieval process and aggregate the collected information, we use the multi-layer perception model (MLP) for state transition modeling,
\begin{equation} \label{eq:state}
    \boldsymbol{s}_{t+1} = \text{MLP}(\boldsymbol{s}_t, [\boldsymbol{u}^{z}_{t+1}, \boldsymbol{i}^{z}_{t+1}]),
\end{equation}
in which $\boldsymbol{u}^{z}_{t+1}$ and $\boldsymbol{i}^{z}_{t+1}$ are the embeddings of the retrieved user and item at time step $t$.

\subsection{ Minimal-sufficient Collaborative Information via Reinforcement Learning}
We follow the standard DDPG \cite{lillicrap2015continuous} reinforcement learning framework to train our retrieval policy with the continuous action space. 
In the Actor-Critic framework \cite{lillicrap2015continuous}, the critic is learning a Q-value function with episodic mini-batch sampled from the replay buffer \cite{mnih2015human},
\begin{equation} \label{eq:critic}
    L(\theta^Q)=\mathbb{E}_{s, a, r, s'}
    \left[r+\gamma Q_{\theta^{Q'}}\left(s', \pi_{\theta^{\mu'}}\left(\cdot | s'\right)\right) - Q_{\theta^Q}\left(s, a\right)\right]^2,
\end{equation}
in which $\theta^{Q'}$ is the target network \cite{lillicrap2015continuous} of the critic, which is fixed during the act network update.
Based on the learning objective in Eq. \eqref{eq:critic}, we can derive the gradient $\nabla_{\theta^Q} L(\theta^Q)$ to update the act network of the critic.
Since the critic provides an approximation of the Q-value function, the optimization step of the actor network can be achieved by policy gradient,
\begin{equation} \label{eq:policy}
    \nabla_{\theta^\mu} L(\theta^\mu)=\mathbb{E}_{s}
    \left[\nabla_a Q_{\theta^Q}\left(s, a\right) \nabla_{\theta^\mu}\pi_{\theta^\mu}\left(\cdot | s\right)  \right],
\end{equation}
similar to the critic network, the policy gradient only updates the act network of the actor, while the target actor network $\pi_{\theta^{\mu'}}$ will be synchronized after each update step.

To further enable continuous space exploration, we follow \cite{lillicrap2015continuous} to add exploration $\mathcal{N}$ noise to the target policy $\pi_{\theta^{\mu'}}$ to find unexplored but informative users and items,
\begin{equation}
    \pi_{\theta^{\mu'}}\left(\cdot | s\right) = \pi_{\theta^{\mu}}\left(\cdot | s\right) + \mathcal{N},
\end{equation}
in which we choose the Ornstein–Uhlenbeck \cite{pavliotis2016stochastic} random process as the exploration process $\mathcal{N}$.

\begin{algorithm}
\begin{algorithmic}
\State \textbf{Input} episode length $L$, Maximum steps in an episode $T$.
\State \textbf{Initialize} actor network $\theta^{\mu}$ and critic network $\theta^{Q}$
\State \textbf{Initialize} target networks $\theta^{\mu'}\leftarrow \theta^{\mu}$ and $\theta^{Q'}\leftarrow \theta^{Q}$ 
\State \textbf{Initialize} the replay buffer $\mathcal{D}=\varnothing$
\While{$l \leq L$}
    \State Receive a user-item pair $z=(u,i)$
    \State \textbf{Initialize} $\mathcal{I}^{\mathit{supp}}_u=\varnothing$, $\mathcal{U}^{\mathit{coll}}_{z}=\varnothing$, $\mathcal{I}^{\mathit{coll}}_{z}=\varnothing$
    \State Construct the prompt of user preference  $\mathcal{I}^{\mathit{supp}}_u$ as in Eq. \eqref{eq:user}
    \State Get the initial prediction $p_0$ according to Eq. \eqref{eq:prp}
    \While{$t \leq T$}
        
        \State \textbf{User-item Retrieval}
        \State Generate the current action $\boldsymbol{a}_{t}$ from the policy $\pi_{\theta^{\mu'}}$
        \State Locate the next user-item pair $(u^{z}_{t},i^{z}_{t})$ as in Eq. \eqref{eq:action} 
        \State Add to the support sets, $\mathcal{U}^{\mathit{coll}}_{z}\leftarrow u^{z}_{t}$ and $\mathcal{I}^{\mathit{coll}}_{z}\leftarrow i^{z}_{t}$ 
        \\
        \State \textbf{Collaborative Prompting}
        \State Construct the prompt of collaborative information $\mathcal{U}^{\mathit{coll}}_{z}$ and $\mathcal{I}^{\mathit{coll}}_{z}$ according to Eq. \eqref{eq:collab}
        \State Get the current prediction $p_t$ as in Eq. \eqref{eq:prob}
        \State Calculate the current reward $r_t$ according to Eq. \eqref{eq:reward}
        \State Observe the next state $\boldsymbol{s}_{t+1}$ according to Eq. \eqref{eq:state}
        \State Store the transition quadruple $(\boldsymbol{s}_t, \boldsymbol{a}_t, r_t, \boldsymbol{s}_{t+1})$ in $\mathcal{D}$
        \\
        \State \textbf{Networks Update}
        \State Sample a minibatch of the quadruple $(\boldsymbol{s}, \boldsymbol{a}, r, \boldsymbol{s}')$ from $\mathcal{D}$
        \State Calculate the minibatch loss $L(\theta^Q)$ for the critic network according to Eq. \eqref{eq:critic}
        \State Update the critic network by the gradient $\nabla_{\theta^Q} L(\theta^Q)$
        \State Update the actor network with the sampled policy gradient according to Eq. \eqref{eq:policy}
        \State Update the target networks:
        \State $\theta^{Q'} \leftarrow \tau \theta^{Q} + (1 - \tau) \theta^{Q'}$
        \State $\theta^{\mu'} \leftarrow \tau \theta^{\mu} + (1 - \tau) \theta^{\mu'}$
    \EndWhile
\EndWhile
\end{algorithmic}
\caption{Training Procedure of CoRAL}
\label{alg:ddpg}
\end{algorithm}

\section{Experiments}
In this section, we conduct extensive experiments on multiple datasets to investigate the following research questions (RQs):
\begin{itemize}
    \item \textbf{RQ1}: How does collaborative information help to align the LLM’s reasoning process to general recommendation tasks?
    \item \textbf{RQ2}: Can CoRAL find sufficient collaborative evidence to enhance LLMs' reasoning?
    \item \textbf{RQ3}: Can CoRAL find minimally-sufficient collaborative evidence to fit the size of prompts?   
\end{itemize}
\subsection{Experimental Settings}

\subsubsection{Datasets} \label{sec:data}
We evaluate CoRAL and baselines on four Amazon Product \cite{ni2019justifying} tasks, which are used in the evaluations of many collaborative filtering methods \cite{zhang2023robust}: 
\begin{itemize}
    \item \textbf{Appliances} refers to a category of home and kitchen devices sold on Amazon. This subset contains 602,777 reviews with 515,650 users and 30,252 products. We use the ``title'' of the items in the metadata as item descriptions.
    \item \textbf{Gift Cards} on Amazon are prepaid store value cards that can be used as an alternative to cash. This subset contains 147,194 reviews with 128,877 users and 1,548 products. We use the ``description'' of the items in the metadata. 
    \item \textbf{Prime Pantry} on Amazon refers to a service offering a wide range of everyday household items and groceries. The subset contains 471,614 reviews with 247,659 users and 10,814 products. We use the original ``description'' in the metadata as the item descriptions.
    \item \textbf{Software} goods on Amazon refer to digital products. The subset contains 459,436 reviews with 375,147 users and 21,663 products. The software product titles in the metadata are used as the item descriptions.
\end{itemize}
Due to the missing descriptions of items in the metadata of some datasets, we use the item titles as the replacement.
To determine the boundary between popular items and long-tail items, we follow the typical 80/20 rule \cite{luke2018recommending,sreepada2020mitigating,yin2012challenging,yuliawati2022long}, which defines the least 80\% items as long-tail items.
Due to the sparsity of the datasets, many users and items only have very few entries in the datasets, in which case collaborative information is almost inaccessible. 
To maintain a certain number of interaction data samples, We follow \cite{zhang2023collm,kim2010collaborative,chen2020efficient,li2021leave} to filter out users and items with fewer than 5 interactions.
For learning-based baselines and CoRAL, we use $70\%$ of the long-tail data and the remaining data in the dataset as the training data.
The remaining $30\%$ of the long-tail data is split equally into the validation and test data. 
We follow the standard preprocessing method \cite{zhang2023collm,wang2023diffusion} to convert the original 5-score into binary labels by the threshold of 3.

\begin{table*}[htp]
\centering
\small
\resizebox{0.75\textwidth}{!}{%
\begin{tabular}{c|cc|cc|cc|cc|cc}
\toprule
 & \multicolumn{2}{c|}{Software} & \multicolumn{2}{c|}{Prime Pantry} & \multicolumn{2}{c|}{Gift Cards} & \multicolumn{2}{c|}{Appliances} & \multicolumn{2}{c}{Average} \\
                           \cmidrule(lr){2-3}  \cmidrule(lr){4-5}   \cmidrule(lr){6-7}  \cmidrule(lr){8-9}  \cmidrule(lr){10-11}  
                                                & AUC     & F1      & AUC     & F1      & AUC     & F1      & AUC     & F1      & AUC      & F1       \\ \toprule
\textbf{AFM} \cite{xiao2017attentional}       & 75.12   & 58.39   & 69.47   & 52.51   & 46.93   & 61.56   & 76.86   & 65.52   & 67.10    & 59.49    \\
\textbf{DCN} \cite{wang2017deep}                & 76.75   & 66.20   & 73.30   & 49.99   & 55.59   & 67.07   & 80.70   & 71.15   & 71.59    & 63.60    \\
\textbf{DFM} \cite{guo2017deepfm}              & 76.04   & 66.63   & 72.92   & 57.86   & 66.76   & 60.01   & 81.83   & 77.37   & 74.39    & 65.47    \\
\textbf{WDL} \cite{cheng2016wide}                 & 78.20   & 69.25   & 73.77   & 56.43   & 60.81   & 57.66   & 73.82   & 74.56   & 71.65    & 64.48    \\ \midrule
\textbf{IPS} \cite{schnabel2016recommendations}       & 78.24        & 71.32        & 72.24        & 61.65        & 64.79        & 63.95        & 82.28        & 75.65        & 74.39         & 66.23         \\
\textbf{CausE} \cite{bonner2018causal}            & 77.78        & 70.84        &  73.69       &  59.80       & 70.51        & 65.39        & 76.86        & 72.04        & 74.71         &  67.02        \\ \midrule
\textbf{LLM-Language} \cite{sanner2023large}     & 73.10   & 66.32   & 51.48   & 41.47   & 83.52   & 74.85   & 74.36   & 70.52   & 70.61    & 63.29    \\
\textbf{CoRAL-random}                            & 77.56   & 58.60   & 64.07   & 50.15   & 91.30   & 59.66   & 77.51   & 61.35   & 77.61    & 57.44    \\ \midrule
\textbf{CoRAL-DFM}                              & 95.25   & 88.68   & \textbf{93.32}   & \textbf{86.73}   & 96.52   & 67.51   & 90.87   & 86.76   & \textbf{93.99}    & 82.42    \\
\textbf{CoRAL-WDL}                              & 93.97   & \textbf{91.18}   & 87.08   & 80.52   & 92.22   & 70.74   & 92.55   & \textbf{89.22}   & 91.45    & 82.92    \\ 
\textbf{CoRAL-AFM}                                  & \textbf{93.99}   & 88.41   & 89.10   & 86.17   & \textbf{98.99}   & \textbf{76.17}   & \textbf{92.66}   & 84.55   & 93.69   & \textbf{83.83}   \\
\textbf{CoRAL-DCN}                              & 91.74   & 87.20   & 85.75   & 77.59   & 97.16   & 70.63   & 91.73   & 86.28   & 91.59   & 80.43   \\\bottomrule
\end{tabular}%
}
\caption{Experimental results (AUC and F1) on four Amazon Product datasets.}
\label{tab:main}
\vspace{-.5cm}
\end{table*}
\subsubsection{Metrics}
We follow the metrics \textit{AUC} and \textit{F1} in long-tail recommendation \cite{zhang2023model,gu2020deep} and collaborative filtering \cite{zhang2023collm,anelli2021reenvisioning}.
The Area Under the Curve (AUC) metric is a performance measurement for classification models that evaluates the tradeoff between true positive rate and false positive rate across different thresholds, where a higher AUC indicates better model performance.
The F1 metric is a statistical measure used in classification tests, combining precision and recall to provide a score that balances both false positives and false negatives, calculated as the mean of precision and recall.

\subsubsection{Baselines}
We introduce baselines in our experiments from three lines of work, collaborative filtering, popularity debiasing, and LLM-based recommendation methods:

Collaborative Filtering:
\begin{itemize}
    \item \textbf{AFM} \cite{xiao2017attentional}: A model learns the significance of each feature interaction from data through a neural attention network. 
    \item \textbf{DCN} \cite{wang2017deep}: A deep neural network featuring a cross-structure is designed for enhanced efficiency in learning specific bounded-degree feature interactions. 
    \item \textbf{DFM} \cite{guo2017deepfm}: A unified neural network architecture for recommender systems is proposed, integrating factorization machines and deep learning. 
    \item \textbf{WDL} \cite{cheng2016wide}: A method that integrates wide linear models with deep neural networks is proposed for enhancing recommender systems. This approach synergistically leverages the strengths of both memorization and generalization.
\end{itemize}

Popularity debiasing baselines directly enhance the collaborative filtering methods via causal debiasing:
\begin{itemize}
    \item \textbf{IPS} \cite{schnabel2016recommendations}: An approach utilizes causal inference techniques to address selection biases in data, ensuring unbiased performance estimation with biased data.
    \item \textbf{CausE} \cite{bonner2018causal}: A domain adaptation technique is developed to train on historical data that captures results from a recommendation system biased by a specific policy and makes predictions for recommendation outcomes under random exposure conditions.
\end{itemize}

To understand the benefit of collaborative information in LLMs, we consider a LLM-based baseline:
\begin{itemize}
    \item \textbf{LLM-Language} \cite{sanner2023large}: A LLM prompting method that describes the user's interacted items and the user's preferences before asking for the user's preference on new items.
\end{itemize}

To understand the behavior of our approach, we consider variants of CoRAL:
\begin{itemize}
    \item \textbf{CoRAL-\textit{Method}}: 
    The collaborative information augmented LLMs, in which the retrieval policy is initialized by the \textit{\textbf{Method}}. The \textit{Method} in our experiments includes DFM, WDL, AFM, and DCN.
    \item \textbf{CoRAL-random}: The LLM is also augmented by collaborative information. However, the retrieval policy is just a rule-based model which randomly retrieves items.
\end{itemize}

\begin{figure*}[htp]
    \begin{subfigure}[b]{0.24\textwidth}
        \centering
        \includegraphics[width=\textwidth]{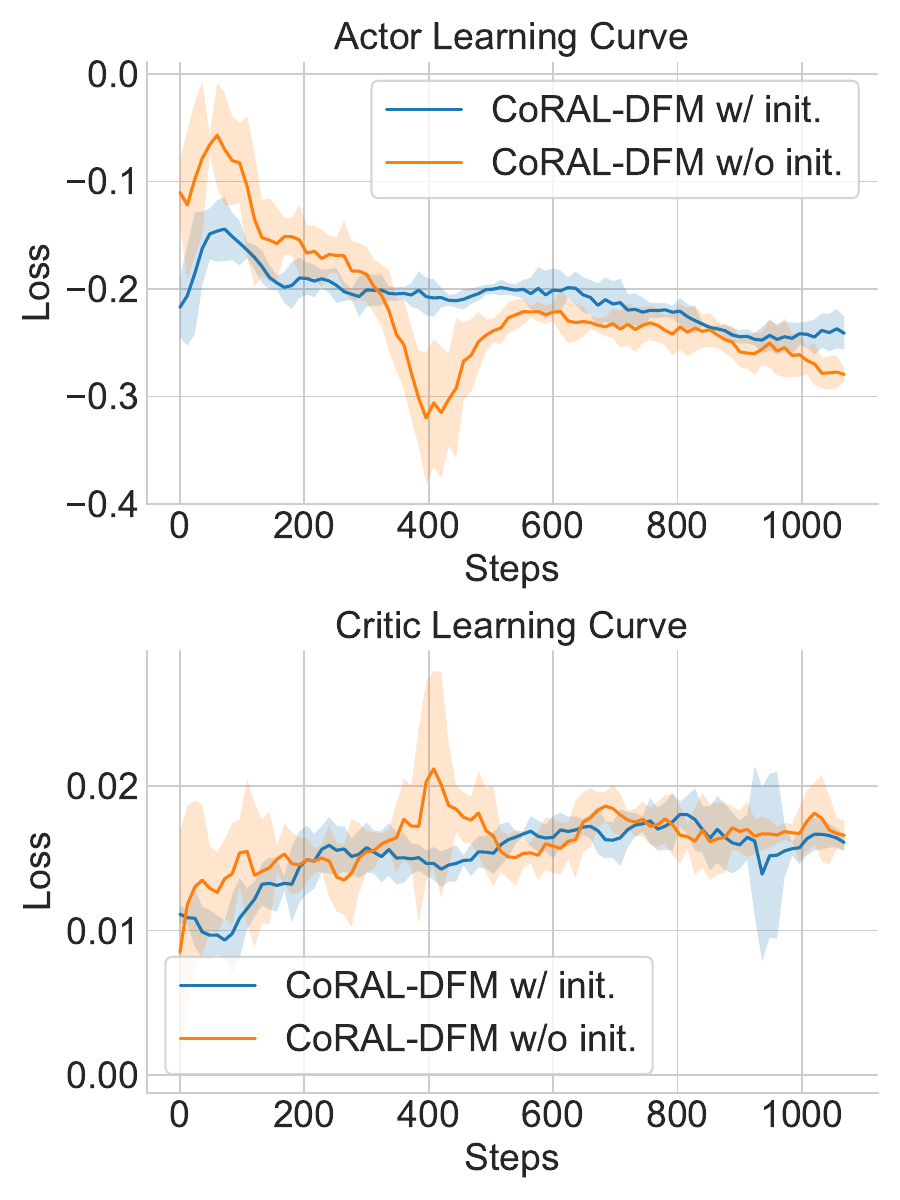}
        \caption{\textbf{CoRAL-DFM} on Gift Cards}
        \label{fig:lc1}
    \end{subfigure}%
    \hfill % Adds horizontal space between subfigures
    \begin{subfigure}[b]{0.24\textwidth}
        \centering
        \includegraphics[width=\textwidth]{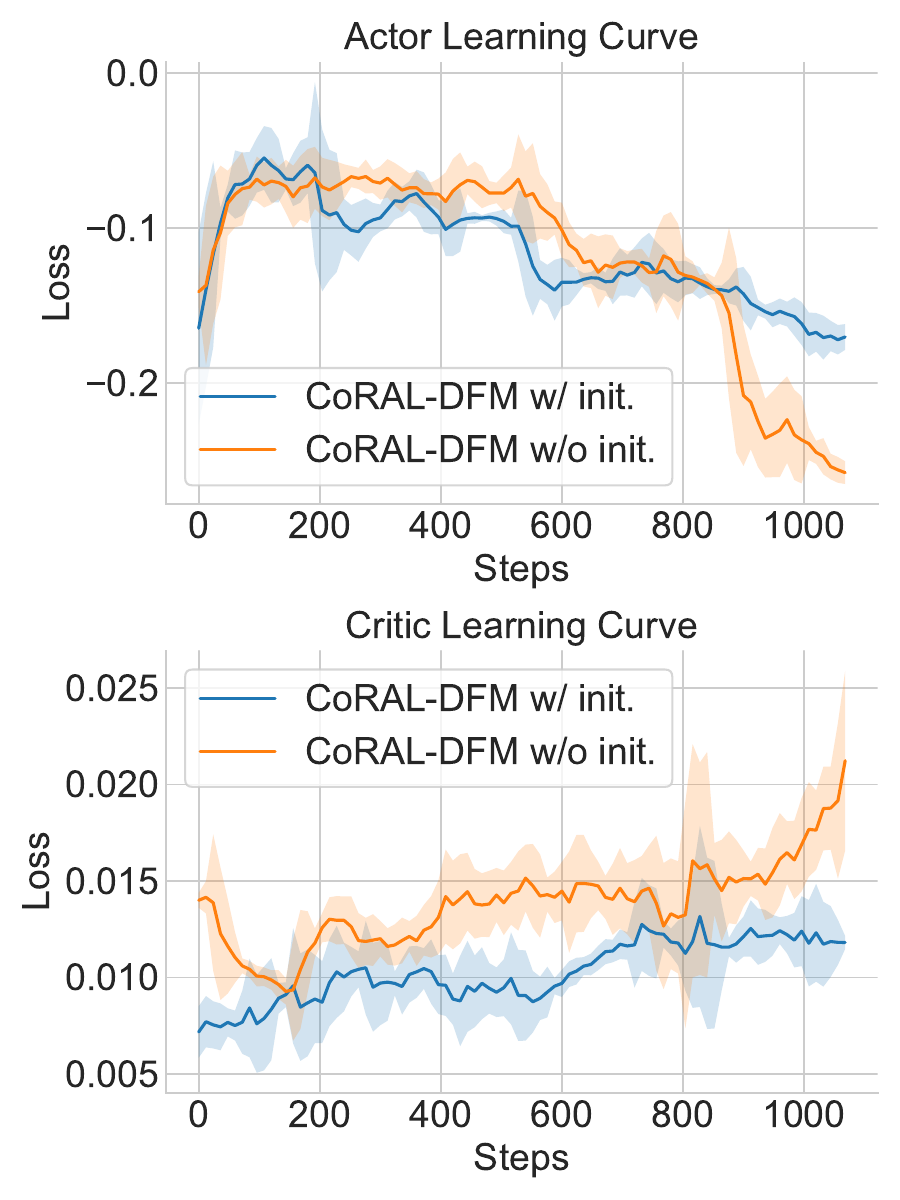}
        \caption{\textbf{CoRAL-DFM} on Prime Pantry}
        \label{fig:lc2}
    \end{subfigure}
        \begin{subfigure}[b]{0.24\textwidth}
        \centering
        \includegraphics[width=\textwidth]{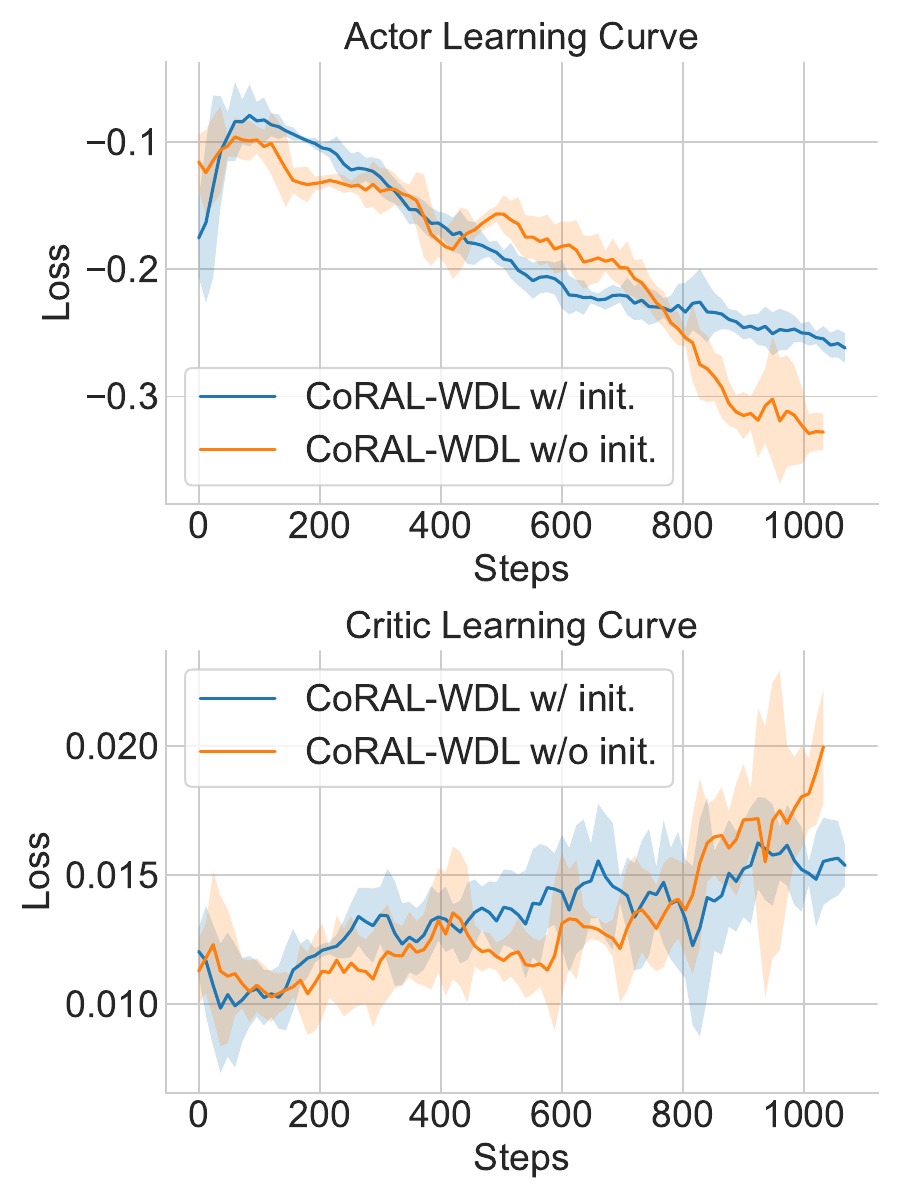}
        \caption{\textbf{CoRAL-WDL} on Gift Cards}
        \label{fig:lc3}
    \end{subfigure}%
    \hfill % Adds horizontal space between subfigures
    \begin{subfigure}[b]{0.24\textwidth}
        \centering
        \includegraphics[width=\textwidth]{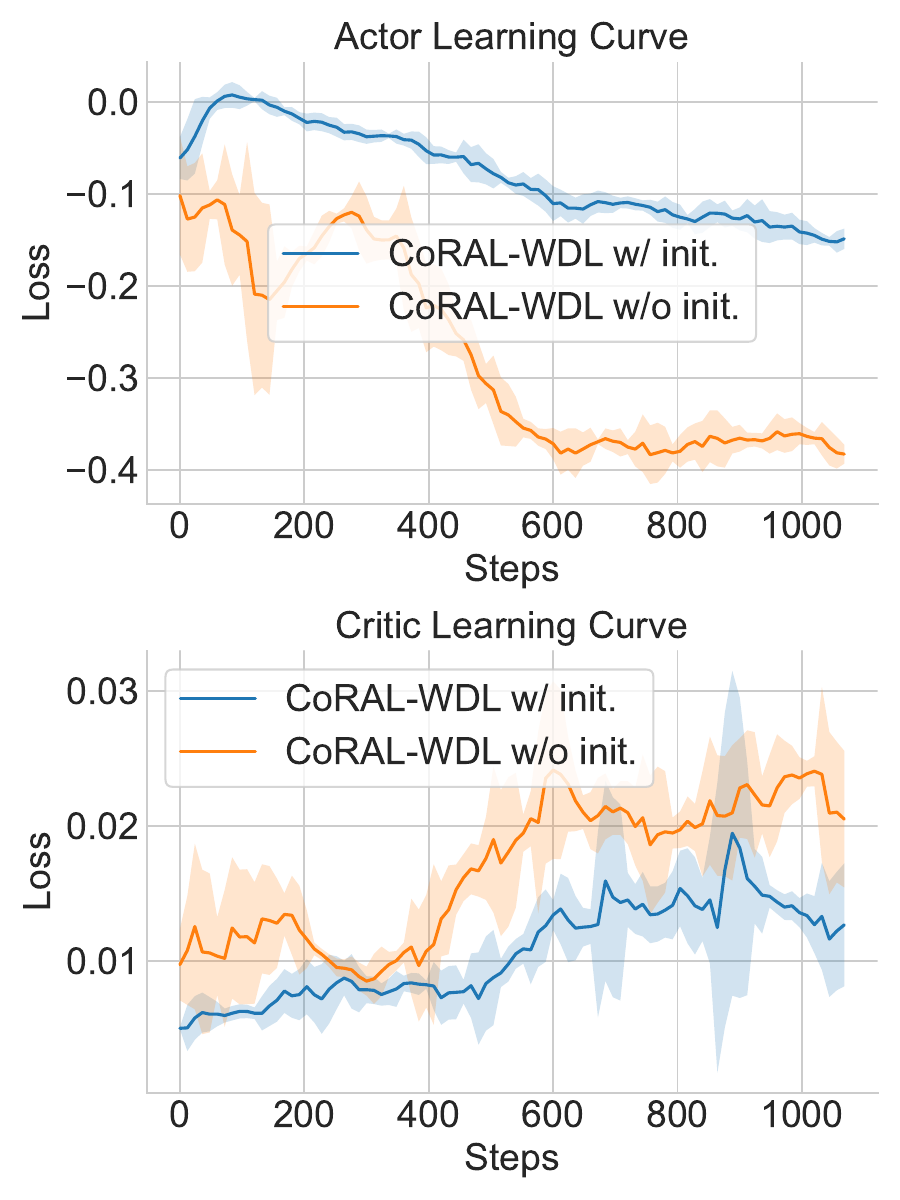}
        \caption{\textbf{CoRAL-WDL} on Prime Pantry}
        \label{fig:lc4}
    \end{subfigure}

    \caption{CoRAL's (DFM and WDL) learning curves on Gift Cards and Prime Pantry datasets.}
    \label{fig:lc}
    % \vspace{-.4cm}
\end{figure*}
\subsubsection{Implementation Details}
We implement our retrieval policy network using PyTorch 2.1. 
For reinforcement learning, the DDPG \cite{lillicrap2015continuous} policy network is implemented using \textit{Stable-Baselines3} \cite{raffin2021stable}.
We set the memory buffer size to $1000$ and the training batch size to $16$ for both the actor and the critic.
For the continuous action for the next user and item, we set the dimensions for both as $128$, which aligns with the size of the user and item embedding.
%as $128$ as well.
The Ornstein–Uhlenbeck noise \cite{pavliotis2016stochastic} added in each dimension of the continuous action space for exploration is set to zero-mean and the standard deviation as $\sigma=0.1$.
The reinforcement learning process starts at the $10$-th iteration, which enables a warm start.
We use Adam optimizer \cite{kingma2014adam} for all the model learning with the learning rate as $0.001$, and we set the maximal learning iterations to $2,000$. 

We implement the reinforcement learning environment using Gym\cite{brockman2016openai}, and we use a GPT-4 \cite{achiam2023gpt} model as the backbone large language model to provide reward.
During the training stage, we allow up to $10$ interactions within a single episode and enable early stop if the absolute value of the discrepancy between the predicted rating and the ground truth rating is less than $0.1$.
During the evaluation stage, for each data sample, we consistently let the policy retrieve $5$ rounds of users and items as collaborative information.

\subsection{Recommendation Performance (RQ1)} \label{sec:exp-main}
We show the comparison results of CoRAL and various baselines to demonstrate the effectiveness of augmenting LLMs with collaborative information as reasoning evidence.

\subsubsection{Effect of the Retrieval Policy.}
In Table \ref{tab:main}, we can observe that by adding collaborative information into the LLM's prompt, 
even if the retrieved users and items are randomly chosen, \textbf{CoRAL-random} consistently outperforms \textbf{LLM-Language} in terms of the AUC scores.
Such observations may imply that collaborative information is still crucial to specific recommendation tasks, even if the LLM can understand the general semantic meanings of the items and the user's preference.
On the other hand, we also observe that \textbf{CoRAL-random} generally performs worse (except for Prime Pastry) than \textbf{LLM-Language} in terms of the F1 scores.
One reasonable explanation is that since \textbf{CoRAL-random} is not specifically curating its selection of users and items, irrelevant information may bring additional bias into the recommendation process and cause the model to have a poor precision-recall trade-off.

\begin{figure*}[htp]
    \begin{subfigure}[b]{0.24\textwidth}
        \centering
        \includegraphics[width=\textwidth]{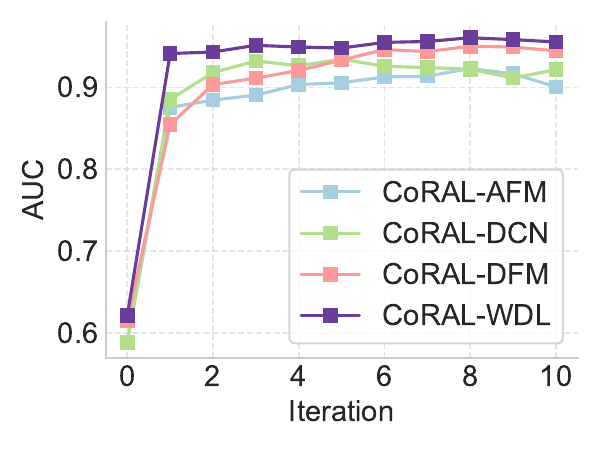}
        \caption{Appliances (AUC)}
        \label{fig:st1}
    \end{subfigure}%
    \hfill % Adds horizontal space between subfigures
    \begin{subfigure}[b]{0.24\textwidth}
        \centering
        \includegraphics[width=\textwidth]{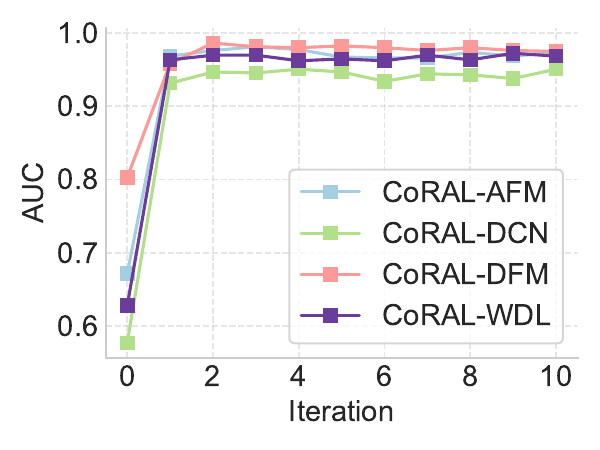}
        \caption{Gift Cards (AUC)}
        \label{fig:st2}
    \end{subfigure}
        \begin{subfigure}[b]{0.24\textwidth}
        \centering
        \includegraphics[width=\textwidth]{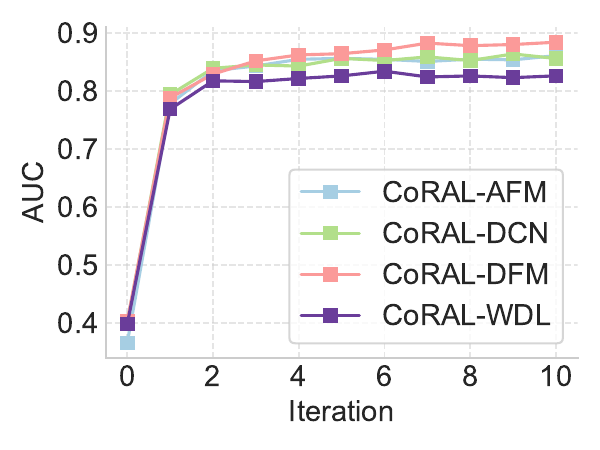}
        \caption{Prime Pantry (AUC)}
        \label{fig:st3}
    \end{subfigure}%
    \hfill % Adds horizontal space between subfigures
    \begin{subfigure}[b]{0.24\textwidth}
        \centering
        \includegraphics[width=\textwidth]{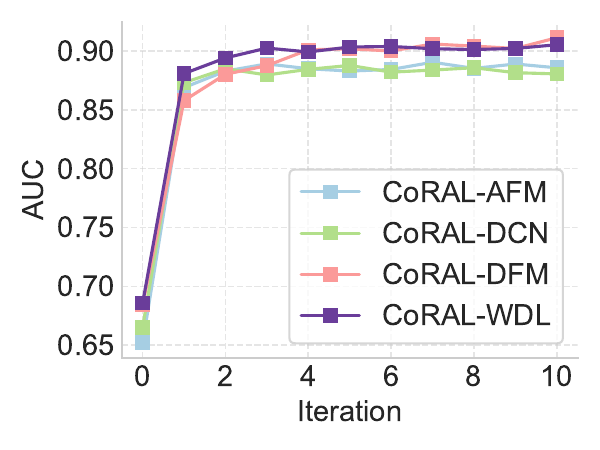}
        \caption{Software (F1)}
        \label{fig:st4}
    \end{subfigure}

    \begin{subfigure}[b]{0.24\textwidth}
        \centering
        \includegraphics[width=\textwidth]{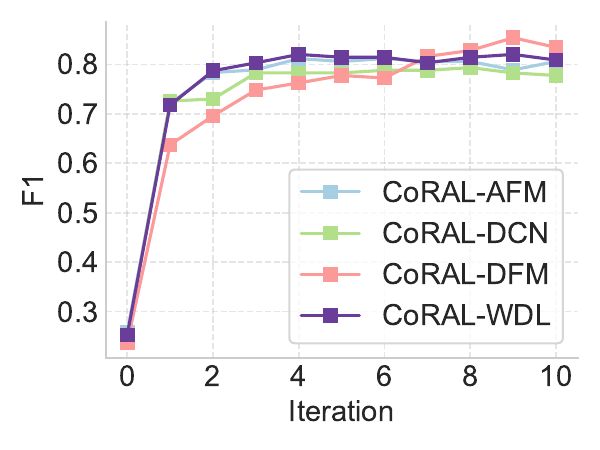}
        \caption{Appliances (F1)}
        \label{fig:st5}
    \end{subfigure}%
    \hfill % Adds horizontal space between subfigures
    \begin{subfigure}[b]{0.24\textwidth}
        \centering
        \includegraphics[width=\textwidth]{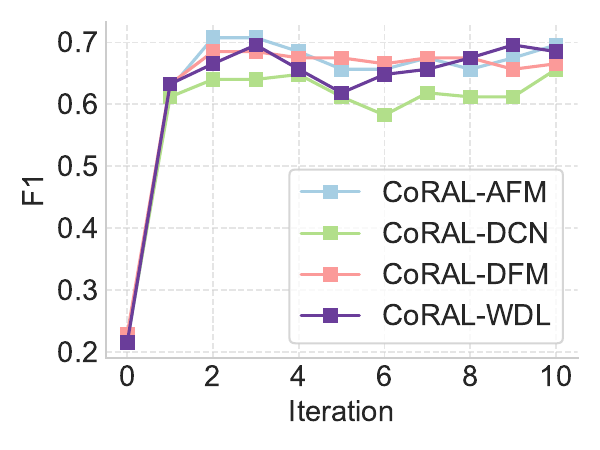}
        \caption{Gift Cards (F1)}
        \label{fig:st6}
    \end{subfigure}
        \begin{subfigure}[b]{0.24\textwidth}
        \centering
        \includegraphics[width=\textwidth]{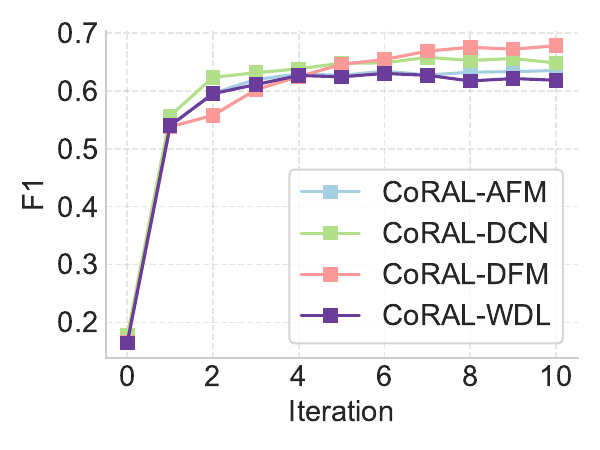}
        \caption{Prime Pantry (F1)}
        \label{fig:st7}
    \end{subfigure}%
    \hfill % Adds horizontal space between subfigures
    \begin{subfigure}[b]{0.24\textwidth}
        \centering
        \includegraphics[width=\textwidth]{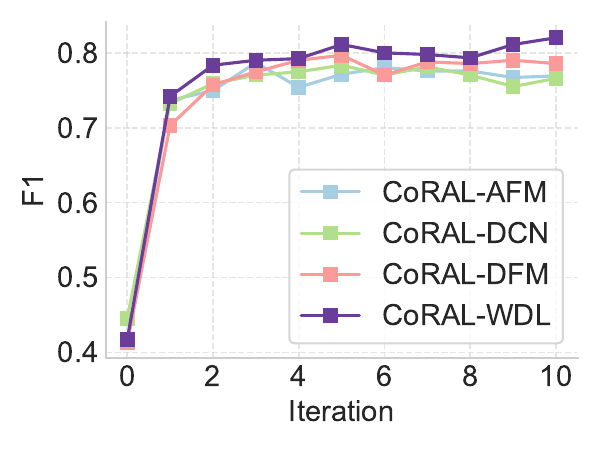}
        \caption{Software (F1)}
        \label{fig:st8}
    \end{subfigure}

    \caption{CoRAL's performance (AUC and F1) w.r.t number of iterations of user-item retrieval on four Amazon Product datasets.}
    \label{fig:ts}
    % \vspace{-.4cm}
\end{figure*}
\subsubsection{Effect of Online Reinforcement Learning.}
In Table \ref{tab:main}, we observe an inconsistency in performance comparison between traditional recommendation baselines and LLM-based baselines, as the LLM-based methods can sometimes (\emph{e.g.}, Prime Pastry and Appliances) perform substantially worse than traditional baselines. This inconsistency in the performance of LLM-based methods suggests the misalignment between the LLM's reasoning and specific recommendation tasks.
The proposed method CoRAL specifically aligns the LLM's reasoning with the recommendation tasks through reinforcement learning. 
With the LLM's reasoning process aligned to the user-item interaction patterns, we can observe significant improvements up to $21.1\%$ and $25.1\%$ for AUC and F1 scores respectively on average.

\subsection{Sufficient Collaborative Information from Popular Items (RQ2)} \label{sec:transfer}
We conduct analytical experiments in this section to show how learning from popular items can benefit online reinforcement learning.
We choose DFM and WDL as the backbone models in the analytical experiments to show their different learning behaviors. 

\subsubsection{Comparison to Randomly Initialized Policy.}
In Table \ref{tab:ablation}, we compare our method, which initializes the retrieval policy by learning from the popular items, with the variant that randomly initializes the policy.
We show that the policies initialized with models learned from popular items are generally performing better than randomly initialized policies, which suggests the data-efficiency advantage of our method.
Since at the early steps of reinforcement learning, the exploration stage may take a long time to navigate and find high-value actions through trial and error,
without efficient exploration strategies or some good embedding spaces, the actor networks could easily overfit and fail to explore better actions.

\subsubsection{Actor-critic Learning Curves}
In Figure \ref{fig:lc}, we show the learning curves of the actor and critic networks, for policies with and without short-head data initialization.
We choose the more challenging datasets, Gift Cards, and Prime Pantry, in terms of methods general F1 performance in Table~\ref{tab:main}, which suggests that these tasks require better balancing between exploration and exploitation.
We can observe a consistent pattern that the actor networks of the policies with random initialization converge faster than policies with short-head data initialization.
However, the critic networks of the policies with random initialization have higher learning loss than policies with short-head data initialization.
Such an observation suggests that the randomly initialized policies could easily overfit and thus the actor-critic learning process can be done asynchronously. 
With the pre-trained user and item embedding spaces on the short-head training dataset, the exploration in the continuous embedding space can be more efficient.
\begin{table}[htp]
\centering
% \small
\begin{tabular}{c|c|cc|cc}
\toprule
                              &     & \multicolumn{2}{c|}{CoRAL-DFM} & \multicolumn{2}{c}{CoRAL-WDL} \\
                              &     & w/ init.      & w/o init.      & w/ init.      & w/o init.     \\ \toprule
\multirow{2}{*}{Software}     & AUC & \textbf{95.25}       & 93.58        & \textbf{93.97}       & 92.35       \\
                              & F1  & \textbf{88.68}       & 88.36        & \textbf{91.18}       & 88.87       \\ \midrule
\multirow{2}{*}{Prime Pantry} & AUC & \textbf{93.32}       & 89.33        & 87.08       & \textbf{89.49}       \\
                              & F1  & \textbf{86.73}       & 80.76        & 80.52       & \textbf{81.86}       \\ \midrule
\multirow{2}{*}{Gift Cards}   & AUC & \textbf{96.52}       & \textbf{96.52}        & 92.22       & \textbf{96.98}       \\
                              & F1  & \textbf{67.51}       & 64.25        & \textbf{70.74}       & 68.81       \\ \midrule
\multirow{2}{*}{Appliances}   & AUC & 90.87       & \textbf{94.48}        & \textbf{92.55}       & 91.84       \\
                              & F1  & 86.76       & \textbf{88.82}        & \textbf{89.22}       & 83.00       \\ \midrule
\multirow{2}{*}{Average}      & AUC & \textbf{93.99}       & 93.48        & 91.45       & \textbf{92.66}       \\
                              & F1  & \textbf{82.42}       & 80.55        & \textbf{82.92}       & 80.64       \\ \bottomrule
\end{tabular}
\caption{Ablation study of CoRAL's performance with or without short-head data initialization for DFM and WDL as the collaborative filtering backbones.}
\label{tab:ablation}
\end{table}

\subsection{Minimally-sufficient Collaborative Information from Iterative Retrieval (RQ3)}
In Figure \ref{fig:ts}, we show the models' performance w.r.t the number of rounds of retrieval. 
We observe that for all the policies of \textbf{CoRAL}, in each iteration, they manage to retrieve informative users and items, while consistently achieving information gain with the information gain margin decreasing.  
Comparing the backbones DFM and DCN of \textbf{CoRAL}, we find a common exploration-exploitation behavior of these two policies, 
as the DCN acts more greedy and reaches its upper-bound performance sooner, while the DFM tends to be more explorative in the early stage and achieves better final performance.
Such an observation suggests the importance of the exploration-exploitation trade-off, which can be more efficiently achieved through the proposed reinforcement learning framework. 
\section{Conclusion}
In this paper, we focus on collaborative filtering-based recommender systems with long-tail items \cite{zhang2023empowering,zhang2021model}.  
We introduce \textbf{CoRAL}, an approach for enhancing long-tail recommendations in traditional collaborative filtering-based recommender systems, overcoming the limitations of data sparsity and imbalance that hamper collaborative filtering methods.
\textbf{CoRAL} integrates collaborative retrieval-augmented LLMs to align the model's reasoning with actual user-item interaction patterns. 
This alignment is pivotal in addressing the common oversight in LLM-based systems that rely heavily on semantic interpretations, neglecting the collaborative dimensions of user-item interactions. 
Additionally, CoRAL employs a reinforcement learning framework to develop a retrieval policy, identifying an optimal set of user-item interactions as the supporting evidence for the LLM's reasoning. 
This strategy ensures minimal yet sufficient collaborative information is used, enhancing the LLM's ability to accurately deduce user preferences and interaction dynamics, hence offering a significant improvement on LLM-based recommendation.

\bibliographystyle{ACM-Reference-Format}
\bibliography{sample-base}

\end{document}